# X-ray linear dichroic ptychography

Yuan Hung Lo[1,2], Jihan Zhou[1], Arjun Rana[1], Drew Morrill[3], Christian Gentry[3], Bjoern Enders[4], Young-Sang Yu[4], Chang-Yu Sun[5], David Shapiro[4], Roger Falcone[4], Henry Kapteyn[3], Margaret Murnane[3], Pupa U. P. A. Gilbert[5], and Jianwei Miao[1†]

[1]*Department of Physics and Astronomy and California NanoSystems Institute, University of California, Los Angeles, CA 90095, USA.*
[2]*Department of Bioengineering, University of California, Los Angeles, CA 90095, USA.*
[3]*JILA and Department of Physics , University of Colorado and National Institute of Standards and Technology (NIST), Boulder, CO 80309, USA.*
[4]*Advanced Light Source, Lawrence Berkeley National Laboratory, Berkeley, CA 94720, USA.*
[5]*Department of Physics, Materials Science and Engineering, Chemistry, and Geoscience, University of Wisconsin, Madison, Wisconsin 53706, USA.*
[†]*Correspondence and requests for materials should be addressed to J.M. (email: miao@physics.ucla.edu)*

**ABSTRACT**

Biominerals such as seashells, corals skeletons, bone, and enamel are optically anisotropic crystalline materials with unique nano- and micro-scale organization that translates into exceptional macroscopic mechanical properties, providing inspiration for engineering new and superior biomimetic structures. Here we use particles of *Seriatopora aculeata* coral skeleton as a model and demonstrate, for the first time, x-ray linear dichroic ptychography. We map the aragonite ($CaCO_3$) crystal *c*-axis orientations in coral skeleton with 35 nm spatial resolution. Linear dichroic phase imaging at the O K-edge energy shows strong polarization-dependent contrast and reveals the presence of both narrow (< 35°) and wide (> 35°) *c*-axis angular spread in sub-micrometer coral particles. These x-ray ptychography results were corroborated using 4D scanning transmission electron nano-diffraction on the same particles. Evidence of co-oriented but disconnected corallite sub-domains indicates jagged crystal boundaries consistent with formation by amorphous nanoparticle attachment. Looking forward, we anticipate that x-ray linear dichroic ptychography can be applied to study nano-crystallites, interfaces, nucleation and mineral growth of optically anisotropic materials with sub-ten nanometers spatial resolution in three dimensions.

# INTRODUCTION

Humans have been using biogenic materials as tools since the dawn of humanity. Biominerals such as bone, teeth, seashells, and coral skeletons exhibit remarkable mechanical properties and complex hierarchical organization (*1*). Due to these unique characteristics, biominerals often outperform their geologic or synthetic inorganic counterparts, thus attracting significant interest in understanding the mechanisms of the biologically-controlled mineralization processes for modern nanotechnology (*2*). Careful understanding of the three-dimensional arrangement of biominerals has important engineering implications, and has led to bioinspired materials that outperform non-biomimetic, inorganic synthetic analogs (*3*).

One of the most common natural biominerals exists in the form of calcium carbonate ($CaCO_3$), which occurs in bacteria, algae, marine organisms and humans (*4*). $CaCO_3$ absorb light anisotropically, such that the π-bonded p orbitals of O and C atoms parallel to the crystal *c*-axis exhibit maximum absorption when aligned parallel to linearly polarized light. The absorption intensity changes with a $cos^2$ law with respect to the azimuthal orientation, and reflects the orientation of the carbonate groups in the crystal. This information can reveal structural and mechanical properties in $CaCO_3$ biominerals (*5*). Coral biomineralization is subject of intense studies, and the mechanisms of crystal nucleation and growth in coral skeletons are only beginning to be revealed (*6*, *7*).

The optical anisotropy in $CaCO_3$ has been leveraged in polarized visible light microscopy to study macroscopic biomineral structure and formation mechanisms (*8*, *9*), and with imaging polarimetry to study crystal orientation uniformity (*10*, *11*). In the shorter wavelength regime, x-ray absorption near-edge structure spectroscopy (XANES) has been used to study the orientations of various polymorphs of calcium carbonates (*12*, *13*), and polarization-dependent imaging contrast (PIC) mapping using x-ray photoemission electron microscopy (X-PEEM) has been demonstrated to quantitatively map crystal orientations in $CaCO_3$ (*13*, *14*). Currently PIC mapping mostly uses X-PEEM in reflection geometry to achieve tens of nanometer resolution. However, PEEM's limited achievable spatial resolution (~20 nm), and the confinement to polished two-dimensional (2D) surfaces are insurmountable limits. Scanning transmission x-ray microscopy (STXM) has taken advantage of dichroic contrast to study polymer fibers (*15*) to resolve 30 nm features, but it is also limited in achievable spatial resolution (~20 nm) by the focusing optics, which have impractical efficiency and working distance.

Although macroscopic morphologies in biominerals have been studied extensively, their nanoscopic structures are still not studied routinely in a quantitative fashion, mostly due to the lack of a proper transmission microscope that offers bulk-sensitive information beyond 5 nm of depth, with spatial resolution better than a few tens of nm. But with the development of high brilliance synchrotron radiation facilities worldwide, advancements in high-resolution imaging techniques, and the increasing availability of insertion device x-ray sources providing polarization control, such as elliptically polarizing undulators (EPU), new synchrotron-based tools are now becoming available for probing nanoscale crystal orientation in $CaCO_3$ minerals and biominerals.

Coherent diffractive imaging (CDI) is such a promising tool for high resolution studies of biominerals (*16*, *17*), which measures the diffraction pattern of a sample and inverts it to a high resolution image using iterative algorithms (*18*). CDI was first applied to image the hierarchical structure of bone at the nanometer scale resolution, revealing the spatial relationship of mineral crystals to collagen matrix at different stages of mineralization (*19*). In particular, ptychography, a powerful scanning CDI technique (*17*, *20*, *21*), has demonstrated 5 nm resolution (*22*) and has attracted significant attention for its general applicability. Ptychography acquires a series of diffraction patterns from spatially overlapping illumination probes on a sample, with phase retrieval algorithms to iteratively recover the incident wave and complex exit wave of the sample. This non-invasive transmission technique offers high resolution imaging of sponge glass fiber (*23*), dentine (*24*), bone (*25*) and frozen-hydrated cells (*26*) in two and three dimensions. Vectorial version of ptychography for studying anisotropic materials has been demonstrated with visible light to study biominerals (*27*), and with x-rays to perform structural and chemical mapping of a meteorite sample (*28*). Synchrotron-based x-ray dichroic ptychography should not only be able to probe materials with the same orientational sensitivity as conventional techniques such as X-PEEM, but can do so with higher spatial resolution, bulk sensitivity, with both absorption and phase information, and in three dimensions.

In this work, we present the first x-ray linear dichroic ptychography of biominerals using the aragonite ($CaCO_3$) coral skeleton of *Seriatopora aculeata* as a test sample. We imaged several coral skeleton particles on and off the O K-edge π* peak and observed significant contrast differences between absorption and phase images. We then performed PIC mapping using three linear dichroic ptychography absorption images to quantitatively determine crystal *c*-axis orientations in the coral with 35 nm spatial resolution. We also qualitatively validated our ptychography results by correlating the ptychography PIC maps with 4D scanning transmission electron microscopy (STEM) (*29*), a scanning nano-electron diffraction technique for probing crystal orientations in crystalline materials. Our results reveal that at the nanoscale, crystallite orientations can be narrowly distributed, as is characteristic of spherulitic crystals, but also randomly distributed in submicron particles. Moreover, we verified, for the first time in any biomineral, linear dichroic phase contrast at a pre-edge energy below the absorption resonance. The use of such phase contrast may lead to new dose-efficient dichroic imaging techniques for studying anisotropic biominerals, and has important implications for understanding the nanoscale organization of crystallites in biominerals.

**MATERIALS AND METHODS**

*Seriatopora aculeata* **skeleton preparation**

The *S. aculeata* coral skeleton used in this study was a pencil-thick, short and tapered branch, termed a nubbin (fig. S1). An entire *S. aculeata* coral skeleton, ~10×10×10 cm$^3$ in size, was purchased from Tropical Fish World, El Cerrito, CA. To remove the tissue and obtain a clean aragonite skeleton the living coral was immersed in 5% sodium hypochloride in water (Chlorox®). After 7 days of bleaching, the skeleton was washed twice in DD-H$_2$O for 5 minutes and twice in ethanol for 5 minutes. A ~1 cm long nubbin, was broken off from the rest of the coral skeleton, placed in an agate mortar and immersed in 100% ethanol, then gently fractured by an agate pestle into micrometer-sized grains. The resulting ethanol suspension was sonicated for 2 minutes for further dispersion, and the supernatant solution was transferred by pipette onto a 200 mesh copper transmission electron microscopy (TEM) grid coated with carbon film and air-dried for 24 hours before data acquisition.

**X-ray linear dichroic ptychography**

Soft X-ray ptychographic microscopy measurements were performed at the imaging branch of the undulator beamline (7.0.1) – COherent Scattering and MICroscopy (COSMIC) – at the Advanced Light Source, Lawrence Berkeley National Laboratory (LBNL) (*30*, *31*). COSMIC provides monochromatic soft x-rays with energies variable from 250 to 2,500 eV, spanning the carbon and sulfur K-edges. Coherent and monochromatic x-rays were focused onto the sample using a Fresnel zone plate, with 45 nm outer zone width, and a total coherent flux of approximately $10^9$ photons/s at the sample position. The TEM grid containing the sample was secured using a Hummingbird 3 mm half-grid tip, then mounted onto a standard FEI CompuStage sample manipulator derived from an FEI CM200 series TEM. Diffraction data were recorded with a 1kFSCCD (*32*), a fast charge-coupled device (CCD) camera developed by LBNL that is capable of 50 frames/sec, 15-bits dynamic range with a 12-bit analog-to-digital converter and with a 1 megapixel detector. Diffraction patterns were acquired without a beamstop and were automatically pre-processed onsite.

Ptychographic measurements consisted of single diffraction patterns recorded at each scan point with 200 or 300 ms dwell time and scanned in a square grid with 40 nm steps to cover an approximately 1.5×1.5 μm field of view, with a reconstruction pixel size of 10.1 nm/pixel. Linear dichroic ptychography data were collected at 0º, 45º, 90º and 135º linear polarizations. The 0º and 90º data were collected with the EPU tuned to horizontal and vertical polarizations, respectively. Since, at the time of these measurements, only linear horizontal and vertical polarizations were under remote computer control, the TEM grid was physically rotated clockwise in-plane by ~135º with respect to the upstream beam, and then the 45º and 135º data were collected using the horizontal and vertical polarizations again.

The same linear dichroic data with 4 polarization angles were collected at two x-ray energies around the O K-edge π* peak: one at pre-edge (534.5 eV, or 1.5 eV before the π* peak) and another at on-peak (536.5 eV, or 0.5 eV after the π* peak maximum). The pre-edge energy used was estimated to be near the negative phase peak and thus had the most negative phase shift relative to vacuum, to achieve optimal phase contrast. The on-peak energy was chosen to be slightly off the maximum absorption peak at 536 eV to reduce beam absorption and attenuation caused by specimen thickness. The resonant energy at the O edge rather than the C edge was chosen in this study because the 3:1 ratio of O to C in $CaCO_3$ means imaging at O resonance gives 3x greater signal-to-noise ratio and contrast. After all data were acquired, ptychography reconstructions were performed using regularized ptychographic iterative engine (rPIE) (*33*) with 300 iterations, with $\beta_{obj} = 0.7$ and $\beta_{probe} = 0.7$, and updating the initial probe only after the 100th iteration.

**X-ray absorption spectroscopy**

Scanning transmission x-ray microscopy with x-ray absorption spectroscopy (STXM-XAS) was measured at ALS beamline 7.0.1.2. The spectromicroscopy data were recorded with 5 ms dwell time and proceeded with 60 nm steps in a square grid scan, with energies spanning the entire O K-edge from 525 to 555 eV. Energy scan steps proceeded with 0.5 eV steps from 525 to 530 eV, then 0.2 eV from 530 to 542 eV, and finally with 0.5 eV steps from 542 to 555 eV. The same energy scan parameters were repeated for x-ray linear polarizations at 0º, 45º, 90º, and 135º. All spectra are normalized via subtraction of the average image from non-resonant energies from 525 to 530 eV.

X-ray absorption spectra were generated using the MANTiS software (*34*). STXM-XAS images at each linear polarization were first converted to optical densities (OD) using fully transmitting regions in the specimen, then aligned using cross correlation. Principal component analysis (PCA) was used to reduce the dimensionality of spectral information in the images to obtain absorption signatures of the coral (fig. S2). The first principal component spectrum at each polarization, which represents the average absorption present in the coral, was shown in this work.

**Ptychographic Polarization-dependent Imaging Contrast (PIC) mapping**

Crystallographic *c*-axis orientations in coral particles were calculated using PIC mapping (*14*), a method that uses linear dichroism effects to quantitatively determine the angular orientation of micro- and nanocrystals. We used the closed-form expression to compute the in-plane angle, $\chi$, and out-of-plane angle, $\gamma$, of the crystal *c*-axis with respect to the linear polarization vector. Here, in-plane is defined as the TEM grid plane that is perpendicular to the x-ray beam. Given three EPU polarization angles 0º, 90º and 45º, the electric field vectors at each polarization is $\vec{E}_1 = E_0\hat{x}$, $\vec{E}_2 = E_0\hat{y}$, and $\vec{E}_3 = (\vec{E}_1 + \vec{E}_2)/\sqrt{2}$, where $\hat{x}$ and $\hat{y}$ are unit vectors. The unit vector describing the *c*-axis orientation is $\hat{c} = \hat{x}\sin\chi\cos\gamma + \hat{y}\sin\chi\sin\gamma + \hat{z}\cos\chi$. For the $i^{\text{th}}$ polarization, the signal intensity is $I_i = I_A + I_B(\vec{E}_i \cdot \hat{c})$, where $I_A$ and $I_B$ are positive fitting parameters. Algebraic manipulations of the three components yield

$$\cos^2\gamma = \frac{1}{2} + \frac{I_1 - I_2}{2[(I_1 - I_2)^2 + (I_1 + I_2 - 2I_3)^2]^{1/2}} \quad (1)$$

$$\sin^2\chi = \frac{1}{I_B}[(I_1 - I_2)^2 + (I_1 + I_2 - 2I_3)^2]^{1/2} \quad (2)$$

Solving for $\gamma$ and $\chi$ in the above equations gives the in-plane and out-of-plane *c*-axis angles, respectively. The range of $\chi$ contracts and expands as $I_B$ varies, but the relative difference in $\chi$ between particles remains consistent. In this work, $I_B$ was arbitrarily set to 2. Since ptychography data at four EPU linear polarizations were collected, two sets of polarizations were used to calculate two PIC maps for each coral particle: the first set used 0º, 45º, and 90º, the second set used 0º, 135º, and 90º.

**PEEM-PIC mapping**

The PIC maps were acquired using the PEEM-3 instrument on beamline 11.0.1.1 at ALS. Nine partly overlapping 60 μm x 60 μm PIC map data were acquired and then tiled and blended in Photoshop®. For each PIC map, a stack of 19 PEEM images were acquired on-peak at the O K-edge π* energy as the linear polarization from the undulator was rotated from 0° to 90° in 5° increments. The 19 images were mounted as a stack and analyzed for fully quantitative crystal orientation information in each 60-nm pixel using the GG Macros in Igor Pro Carbon®. As coral skeletons are made of <99.9% aragonite ($CaCO_3$) and <0.1% organic matrix (*35*), the contribution of organics to oxygen spectroscopy is <<0.1%, which is not expected to exhibit any polarization dependence. Thus, PIC mapping in ptychography or PEEM only displays aragonite crystal orientations.

**4D-STEM and electron tomography**

Scanning nano-diffraction (4D-STEM) data and electron tomography data were collected at the National Center for Electron Microscopy, Molecular Foundry, LBNL. Both methods were used on precisely the same three particles already analyzed with ptychography. A Titan 60-300 equipped with an Orius 830 detector (Gatan, Pleasanton, CA) and four windowless silicon drift EDS detectors (FEI super-X) were used with a solid angle of 0.7 srad. The microscope operated in STEM mode at 200 kV with an electron beam current of ~16 pA for 4D-STEM datasets and ~40 pA for STEM imaging. The 4D-STEM diffraction data were taken on Orius CCD with a camera length of 300 mm using a convergence angle ~0.51mrad, with 64 × 64 square grid scan positions. Before clustering of 4D-STEM data, individual diffraction patterns were preprocessed by aligning the center of mass of the main beam to the image center to correct for horizontal and vertical shifts introduced by beam tilt.

A diffraction similarity map was generated using agglomerative hierarchical clustering (*36*) of 4D-STEM data. Agglomerative hierarchical clustering initializes all data points, or individual diffraction patterns, as independent clusters. The algorithm then computed the proximity between every pair of data points using a specified distance metric (e.g. Euclidean distance, cosine similarity, correlation). Next, pairs of data points were linked to one another using a specified linkage metric (e.g. average distance, centroid distance, nearest neighbor distance) to form new grouped clusters, and repeat until all data points were linked together into a hierarchical tree. Finally, the consistency of the resulting clusters was verified by evaluating the distances between each pair of neighboring clusters in the tree. A distance that is greater than a predefined inconsistency score constituted a natural partition between clusters, such that separate clusters were considered to be truly independent. This clustering was performed in MATLAB (MathWorks) environment with the "linkage" function, using correlation as the distance metric, nearest neighbor as the linkage metric, and an inconsistency score of 1.2.

Electron tomography was performed using the GENeralized Fourier Iterative REconstruction (GENFIRE) (*37*), an algorithm that has been used to determine the 3D and 4D atomic structure in materials with unprecedented detail (*38–40*). Before reconstruction, STEM projections were aligned to a common tilt axis using the center of mass and common line methods (*41*). Next, a constant background – the average value in an empty region of the image – was subtracted from each projection, and the process was optimized by minimizing the differences between all common lines and a reference common line. the projections were then normalized to have the same total sum, since the integrated 3D density of the isolated coral particle should be consistent across all tilt angles. The preprocessed projections were used in GENFIRE reconstruction, which ran for 100 iterations with oversampling ratio of 2 (*42*), 0.7-pixel interpolation distance, and the enforcement of positivity and support constraints.

**RESULTS**

Figure 1 shows the experimental schematic of linear dichroic x-ray ptychography experiment. Fig. 1C shows the O K-edge spectra obtained from STXM-XAS at each linear polarization, showing the expected dependence of x-ray absorption on the relative angle between crystal *c*-axis and x-ray polarization (*13*). The π* peak absorption occurs around 536 eV and is maximum when x-ray polarization is parallel to the π orbitals of C and O in the trigonal planar carbonate group. The broad σ* peak occurs around 547 eV and is anticorrelated with the π* peak.

To study the effects of linear dichroism on the absorptive component of the coral's complex exit wave, we imaged 3 coral particles at two energies, pre- and on-peak. Figure 2A shows on-peak ptychography absorption contrast images of 3 coral particles at 0º, 45º, 90º, and 135º polarizations from top to bottom, and from left to right the three particles are denoted as P1, P2 and P3, respectively. Relative contrast within the particles changes dramatically with polarization, signifying the presence of differently oriented nanoscale domains in each particle. P1 displayed overall smooth features with little internal structures, whereas P2 and P3 contained multiple nano-domains and striations. While resonant imaging revealed rich polarization-dependent absorption contrast due to the linear dichroism, imaging off resonance produced no absorption contrast when varying polarizations (fig. S3A).

To examine the effects of linear dichroism on the phase component of the coral skeleton particles' complex exit wave, we also collected ptychography images at 534.5 eV (Fig. 2B), an energy slightly before the π* peak. In general, the negative phase peak is at lower energy than both the positive phase peak and the absorption peak and provides the greatest contrast with respect to non-resonant material (*43*). Phase contrast images reveal sharp boundaries and complex surface morphologies in the particles. On the other hand, on-edge phase images of the coral particles reveal polarization-dependent contrast that agree very well with on-edge absorption images (fig. S3B). According to the Kramers-Kronig relation, the effects of linear dichroism in crystal orientation manifests in both components of the complex refractive index. But as we observe in ptychography maps, while the effect on absorption is significant on resonance (Fig. 2A), the effect on phase is maximum off resonance (Fig. 2B). Resolution of the ptychography images is estimated to be 35 nm using the knife-edge method with 10-90% intensity cutoff (fig. S4). Given a mass attenuation coefficient ($\mu/\rho$) of $2 \times 10^4$ cm$^2$/g for CaCO$_3$ on O K-edge and a ~50% overlap, each recorded projection absorbed an estimated dose of $1.44 \times 10^8$ Gy. At this dose and estimated resolution no noticeable deterioration was observed in the sample (*44*).

Ptychographic PIC mapping revealed that the orientations of crystals are much more diverse at the nanoscale than previously appreciated. As is clear from Fig. 3, and in particular from the broad range of colors in all 3 particles in Fig. 3A, and the large width of the histograms in Fig. 3B, many crystallites are present in what was previously assumed to be single crystals, e.g. P2, or two crystals, e.g. P1. These crystallites vary in orientation gradually, as displayed by color gradients across all larger domains, as in the mustard color domain of P1, or the green-blue domain of P2, or the red-blue domain of P3. There are also unexpected, smaller (~100 nm) domains with orientation different from the larger domain, but not randomly oriented as expected from sample preparation artifacts, e.g. randomly aggregated particles. These small domains are co-oriented with one another but spatially separate from one another. See for example in P1 the string of smaller green crystallites at the bottom of the mustard crystal, or the four green crystallites on the right side, or the smaller blue domains within the red domain. Other smaller domains in P2 are the red-dot crystallites near the edges. P3 shows several blue-green crystals of similar smaller sizes and orientations that are interspersed with the rest of the particle, which is shown as red-yellow. These nano-crystallites are highly surprising, as they were not revealed by previous methods, such as PIC mapping using X-PEEM in Fig. 4. PEEM-PIC mapping shows that the smaller crystalline domains in the centers of calcification (CoCs) are randomly oriented (Fig. 4). Ptychography PIC mapping, instead, shows that several smaller (~100 nm) crystallites are mis-oriented with respect to the larger crystals domains in which they are embedded, but they are co-oriented with one another (Fig. 3A).

To better understand the co-oriented smaller domains, we performed PIC mapping on the on-edge linear dichroic ptychography absorption images to quantitatively map $c$-axis angles in the coral particles and analyzed the in-plane ($\gamma$) and out-of-plane ($\chi$) $c$-axis orientation angles. Since Eq. 1 and Eq. 2 only require 3 polarizations to compute the $\gamma$ and $\chi$, PIC maps in Fig. 3A were calculated using the 0º, 45º and 90º polarization images, and a second set of PIC maps were computed using 0º, 135º and 90º polarization images with consistent results (fig. S5). In each PIC map, in-plane angles are color-coded according to the coral's crystal axes relative to the x-ray polarization, which is horizontal at 0º. Orientation ranges from 0º to 90º, since angles beyond that range are degenerate and cannot be distinguished from contrast alone. The out-of-plane angles between the $c$-axes and x-ray polarization are represented by brightness, such that $c$-axes aligned with the imaging plane are displayed with high brightness, and $c$-axes that are perpendicular to the imaging plane are shown in low brightness, and is lowest when the axes align directly with the x-ray beam.

Histograms of $\gamma$ (top) and $\chi$ (bottom) in Fig. 3B present the *c*-axis angular distribution derived from the PIC maps. P1 exhibits two distinct subdomains, within which the angular spread is <35°, but these are oriented more than 35° apart from one another. In contrast, P2 and P3 show greater sub-micrometer orientational fluctuations that span more than 35°, suggesting that particles P2 and P3 comprise many differently oriented nanocrystals from CoCs. To further examine the abrupt orientational change between subdomains in P1, we performed STEM tomography on the very same P1 particle, which reveal two separate particles on top of each other and thus confirmed the ptychography results (fig. S6).

To further validate the localization and orientation of crystallites observed in ptychography PIC maps, we collected scanning electron nano-diffraction 4D-STEM data on particle P3 and assessed its nanoscale lattice changes over the entire particle. The converging beam electron diffraction (CBED) patterns were analyzed using unsupervised agglomerative hierarchical clustering (*36*) to sort the particle into different regions with similar crystal orientations. Fig. 5A shows a STEM image of P3, and Fig. 5B shows the resulting similarity ranking map generated by hierarchical clustering. The closer the regions are in color, the more similar their corresponding CBED patterns are. Representative CBED patterns from the coral are shown in Fig. 5C, with numbers corresponding to the labelled regions in Fig. 5B.

The CBED patterns reveal variations and similarity in diffraction – hence crystal orientations – across the particle. For instance, pattern #1 is similar to pattern #9, and both are in close proximity to one another. In contrast, patterns #4 and #6, although within the same region, have dissimilar diffraction patterns. Moreover, the similarity ranking map divides the particle P3 into sub-regions closely resembling those shown in the ptychography PIC map in Fig. 3A. In particular, distinct subdomains in regions #2, #3, #4 and #7 of the similarity ranking map match well with the corresponding areas in the ptychographic PIC map. This result serves as further confirmation of the orientation heterogeneity within P3.

**DISCUSSION**

X-ray linear dichroic ptychography of coral particles shown in Fig. 2A and B unveil strong polarization-dependent absorption and phase contrast that is evidence of differently oriented subdomains in each particle. Moreover, each particle exhibits diverse structural features and contains crystal orientation domains that range in size from tens to hundreds of nanometers. While both on-edge absorption and pre-edge phase images reveal fine internal features in the coral skeleton particles, phase images seem to be more sensitive to edges and thus show surface morphologies and boundaries more clearly. The use of phase information to visualize weakly scattering fine features has previously been demonstrated with visible light phase ptychography to enhance cellular contrast in live cells (*43*, *45*). In the case of biominerals, the simultaneous phase and absorption contrast imaging provided by x-ray ptychography can be used to probe nanoscale boundary features beyond the surface, enabling structural study of inter-crystal topology that is critical in understanding biomineral nucleation and growth. To the best of our knowledge, this is the first demonstration of combined linear dichroic absorption and phase imaging of optically anisotropic materials.

The orientations observed in the main domains of each particle are <35°, as previously observed by Benzerara et al., Sun et al., and Coronado et al. (*6*, *46*, *47*), and is fully consistent with spherulitic crystals (*6*). PIC maps generated from the x-ray linear dichroic ptychography images (Fig. 2 and fig. S3) provide quantitative crystal orientation information with high resolution, and at a depth on the order of 100-500 nm, which is not available with the 5-nm-surface-sensitive X-PEEM PIC mapping (Fig. 4). At a fine-grain level, the ptychography PIC map of P1 shows the presence of two overlaid homogeneous particles, each having a *c*-axis angular spread <35° (Fig. 3B). Such narrow angular spread is typical of spherulitic crystal such as those that form the all coral skeletons, fills space isotropically with anisotropic crystals, and thus provides the coral skeleton with the needed structural support (*6*).

Similarly, the main crystalline domains in all three particles P1, P2, and P3 are co-oriented within 35°, as expected from spherulitic crystals. Unexpectedly, all three particles exhibit several smaller (~100 nm) domains differently oriented with respect to the main domain (Fig. 3A and 3B). Since the orientations of these smaller domains are not random, but co-oriented with one another, these crystallites cannot be the randomly oriented nanocrystals observed in the CoCs in Fig. 4. Note that these smaller domains in corals have not been observed with such detail before, presumably because previous studies did not have the capability to detect bulk subdomain morphology. Are these co-oriented but disconnected corallites consistent with either of the current models proposed for coral skeleton formation? In one model corals form their skeletons by either ion-by-ion precipitation from solution (*48*, *49*) in the other by attachment of amorphous precursor particles (*7*). Neither model explicitly predicts the nucleation and growth of co-oriented, disconnected crystals.

The latter are unlikely to result from extraneous contaminants during coral skeleton growth, or depositions after the death of the animal, or aggregation of particles after fracturing, as all three scenarios would not generate either the co-oriented or the equally sized crystallites consistently observed in all three particles. Although additional evidence is needed to fully understand the source and formation mechanism of the observed co-oriented, disconnected crystallites, any model for coral skeleton formation must be consistent with their formation.

One parsimonious explanation is that the three particles analyzed were fractured from interfacial regions, at the boundary of two main crystal orientation domains, for instance two adjacent fibers. Assuming a non-straight, jagged boundary between two crystal fibers, what appears as smaller separate crystallites may be portions of the same larger crystal. This parsimonious interpretation does not require a new understanding of coral formation; it is consistent with coral formation by attachment of amorphous nanoparticles (*7*), and crystallinity propagating through the amorphous phase one particle at a time (*50*). In this crystallinity originating from two distinct nucleation events would propagate in space until the two crystal orientations abut one another. Since they crystallized one particle at a time, their boundary between the two adjacent crystal fibers is not a straight plane but a nanoparticulate surface. A cross-section of such surface would appear as a jagged edge in a 2D map. The PIC map of the pristine coral skeleton geometry in Fig. 4 does indeed show jagged edges between fibers, and therefore supports this interpretation. The dashed line in Fig. 4C is a hypothetical fracture line at the jagged interface of two fibers, which would result in disconnected but co-oriented crystallites. Thus, the observation of the co-oriented smaller crystallites demonstrates the potential of ptychographic PIC map in testing hypotheses for coral skeleton nucleation and growth.

To put ptychographic PIC map's fine-grain, nanoscale results into the larger perspective, we also used X-PEEM to acquire a wide-field of view, lower spatial resolution (60 nm) PIC map from another region of the same *S. aculeata* skeleton (Fig. 4). The X-PEEM PIC map reveals two main types of crystallites on a macroscopic level. One type consists of large, micrometer size spherulitic crystals with less than 35° angular spread, which resembles P1 and its two narrowly distributed orientations. Another type consists of submicron crystals with randomly oriented *c*-axes, localized in the CoCs (Fig. 4), which agrees with P2 and P3 and their more broadly distributed angular spreads (Fig. 3B). By correlating nanoscopic ptychography PIC map with its microscopic X-PEEM analog, we get a multi-length scale picture of the coral skeleton architecture. The co-oriented, disconnected crystallites, however, are not detected by PEEM-PIC mapping, because its maximum probing depth is 5 nm at the O K-edge (*51*).

Ptychographic PIC mapping is qualitatively validated with 4D-STEM (Fig. 5B), in which hierarchical clustering sort the similarity of CBED patterns into a hierarchical tree. Regions with similar CBED patterns can be assumed to share the same crystal orientations. Comparison between ptychography PIC map of P3 in Fig. 3A and similarity rank map in Fig. 5B show mostly consistent subdomains. While it is possible to simulate CBED patterns using a known aragonite model at various orientations and match them with experimental patterns to estimate *c*-axis orientation (*52*), the coral particle's arbitrary thickness near the center attenuated much of the beam, thereby making exact comparison difficult. In the future, more careful sample preparation using focused ion beam can produce coral specimens with desired thickness for optimal x-ray and electron transmission. Nonetheless, this work demonstrates a new attempt to correlate x-ray ptychography with 4D-STEM to understand nanoscale crystal orientations in biomaterials.

There are a few limitations in this work. First, since the aragonite particles studied here are randomly oriented polycrystals, x-ray linearly dichroic ptychography images actually measured the integrated sum of all c-axes along the beam direction, with contributions to the contrast from different nano-crystals. Therefore, PIC maps presented here represent only the average orientations in the coral particles. One strategy to overcome this limitation may be to use a ptychographic vector nanotomography approach similar to the ones used to study 3D magnetic vector field (*53*).

By doing such vector tomography reconstruction, one should be able to obtain true *c*-axis orientations of voxels in three dimensions. Second, since the coral specimens imaged here were ground by a pestle and thus produced uneven surface, the uncontrolled sample morphology makes quantitative interpretation of ptychographic phase images more complicated, as linear dichroic phase contrast then becomes a function of both specimen thickness and crystal orientation.

Also, the process of physically grinding the coral skeleton fractured the skeleton into particles and may have introduced structural artifacts. This can be alleviated by first embedding the specimen in epoxy resin then using focused ion beam (FIB) to make it suitably thin for ptychography. Another limitation of the current work is that for simplicity we considered the refractive index of aragonite to be a scalar, not a tensor. For fully quantitative x-ray ptychography all calculations will have to be re-done considering the tensor refractive index of aragonite (*54*). Although the observation of linear ptychographic phase dichroism is interesting, more careful sample preparation and measurements are needed in future studies to extract quantitative *c*-axis orientation information from phase.

The x-ray linear dichroic ptychography results presented in this work imply an important possibility. Conventionally, enhanced polarization-dependent contrast is derived from absorption contrast when imaged on elemental absorption edges, with the trade-off that more energy is deposited into the sample per unit area and unit time, which inevitably exacerbates sample radiation damage. However, as this work has demonstrated, one major benefit of x-ray linear dichroic ptychography is that strong polarization-dependent phase contrast is also available when imaged off of resonant energy. This suggests that linear dichroic phase contrast imaging offers an alternative path to obtaining quantitative crystal orientation insights without having to subject the sample to the same radiation dose as absorption edge imaging. This potentially important finding can enable more sophisticated and data-intensive studies. Such dose-efficient technique will be especially advantageous when acquiring vector tomography datasets, since many tilt projections are needed to achieve high quality 3D reconstruction. So far most linear dichroism studies focused on absorption as linear phase dichroism is difficult to obtain experimentally. Therefore, the x-ray linear dichroic phase ptychography technique presented here has the potential to become an important tool for studying dose-sensitive materials.

In the future, x-ray linear dichroic ptychography can be applied to image other materials such as tooth enamel, bone, seashells, sea urchin spines, polymers, and other classes of optically anisotropic materials with sophisticated nanoscale morphologies and crystallinity. It can also be used to study intricate mechanisms such as crystal nucleation, self-assembly, phase transitions and space-filling growth. In addition, given the ease with which x-ray linear polarization can be tuned at synchrotrons, such technique can be readily implemented in existing coherent diffractive imaging beamlines, and can be combined with x-ray absorption spectroscopy and x-ray fluorescence imaging to produce multidimensional images of heterogeneous samples. As coherent flux, data acquisition, and big data technologies continue to advance at synchrotron radiation facilities, we envision this linear dichroic ptychography imaging technique becoming part of a powerful suite of tools that offer both compositional and orientational information in real time.

**CONCLUSION**

In this work we present the first demonstration of x-ray linear dichroic ptychography. By imaging three *S. aculeata* coral skeleton particles at pre- and on-peak at the O K-edge, we observed strong polarization-dependent phase and absorption contrasts. Then we performed PIC mapping on the dichroic ptychography absorption images to quantitatively estimated *c*-axis orientations in the corals and observed the presence of two types of previously known crystallites: the main domain in each particle with the narrow angular spread < 35° consistent with spherulitic crystals, and with the randomly oriented sub-micrometer nanocrystal domains observed in the centers of calcification in *Acropora* skeletons. Finally, we observed ~100 nm crystallites mis-oriented with respect to the main domains in each particle but co-oriented with one another, disconnected, interspersed with and within larger crystals which are consistent with coral skeleton formation by attachment of amorphous particles, which crystallize with jagged edges in both coral fibers and CoCs.

We also validated the x-ray results with 4D-STEM and confirmed that regions of orientational diversity are largely consistent. The observation of strong linear phase dichroism off of absorption edge offers the potentially interesting possibility of using phase imaging rather than absorption imaging in future linear dichroism studies as a way to alleviate sample radiation damage. With further development, we expect x-ray linear dichroic ptychography to become a powerful non-destructive tool for probing general classes of optically anisotropic materials such as biominerals with sub-ten nanometer resolution in two and three dimensions.

**ACKNOWLEDGEMENT**

We thank Marcus Gallagher-Jones for his guidance on 4D-STEM analysis, and Jared J. Lodico and Billy A. Hubbard for their help handling the samples during the ptychography experiments at COSMIC. We also gratefully acknowledge the support of NVIDIA Corporation for the donation of the Quadro K5200 GPU used for this research. **Funding**: This work was primarily supported by STROBE: A National Science Foundation Science & Technology Center under Grant No. DMR 1548924. J.M. acknowledges partial support of the Department of Energy (DE-SC0010378) for the development of GENFIRE. P.G. gratefully acknowledges 80% support from DOE-BES-CSGB-Geosciences grant DE-FG02-07ER15899, and 20% support from NSF-BMAT grant DMR-1603192. All x-ray experiments were done at the Advanced Light Source, which is supported by the Director, Office of Science, Office of Basic Energy Sciences, US Department of Energy under Contract No. DE-AC02-05CH11231. The 4D-STEM and electron tomography experiments were performed at the Molecular Foundry, which is supported by the Office of Science, Office of Basic Energy Sciences of the U.S. DOE under Contract No. DE-AC02-05CH11231. The electron tomography data was based on high-angle annular dark-field STEM images. D.M. acknowledges support by the Department of Energy National Nuclear Security Administration Stewardship Science Graduate Fellowship program, which is provided under grant number DENA0003864. **Author contributions**: J.M. directed the project; Y.H.L., J.M., P.G., C.Y.S., D.S., M.M., and H.K. planned the experiments; P.G. provided the samples; J.Z. collected the 4D-STEM and electron tomography data; Y.H.L., A.R., C.G., D.M. and D.S. collected the ptychography and STXM data; Y.H.L. performed the tomography reconstructions with input from J.M.; Y.H.L. performed PIC map analysis; P.G. and C.-Y.S. acquired and processed the PEEM-PIC maps; All authors contributed to the discussion and writing of the manuscript. **Author Affiliations**: H.K. is partially employed by KMLabs Inc. **Competing Interests**: The authors declare that they have no competing interests. **Data and materials availability**: X-ray linear dichroic ptychography data presented in this work are available for download at CXIDB, and electron tomography data presented in this work are available for download at Mendeley Data. All data needed to evaluate the conclusions in the paper are present in the paper and/or the Supplementary Materials. Additional data related to this paper may be requested from the authors.

**REFERENCES**
1. H. A. Lowenstam, S. Weiner, *On Biomineralization* (Oxford University Press, 1989).

2. U. G. K. Wegst, H. Bai, E. Saiz, A. P. Tomsia, R. O. Ritchie, Bioinspired structural materials. *Nat. Mater.* **14**, 23–36 (2015).

3. Z. Yin, F. Hannard, F. Barthelat, Impact-resistant nacre-like transparent materials. *Science*. **364**, 1260–1263 (2019).

4. N. K. Dhami, M. S. Reddy, A. Mukherjee, Biomineralization of calcium carbonates and their engineered applications: a review. *Front. Microbiol.* **4** (2013), doi:10.3389/fmicb.2013.00314.

5. J. Stöhr, K. Baberschke, R. Jaeger, R. Treichler, S. Brennan, Orientation of Chemisorbed Molecules from Surface-Absorption Fine-Structure Measurements: CO and NO on Ni(100). *Phys. Rev. Lett.* **47**, 381–384 (1981).

6. C.-Y. Sun, M. A. Marcus, M. J. Frazier, A. J. Giuffre, T. Mass, P. U. P. A. Gilbert, Spherulitic Growth of Coral Skeletons and Synthetic Aragonite: Nature's Three-Dimensional Printing. *ACS Nano*. **11**, 6612–6622 (2017).

7. T. Mass, A. J. Giuffre, C.-Y. Sun, C. A. Stifler, M. J. Frazier, M. Neder, N. Tamura, C. V. Stan, M. A. Marcus, P. U. P. A. Gilbert, Amorphous calcium carbonate particles form coral skeletons. *Proc. Natl. Acad. Sci.* **114**, E7670–E7678 (2017).

8. M. J. Olszta, D. J. Odom, E. P. Douglas, L. B. Gower, A New Paradigm for Biomineral Formation: Mineralization via an Amorphous Liquid-Phase Precursor. *Connect. Tissue Res.* **44**, 326–334 (2003).

9. I. M. Weiss, N. Tuross, L. Addadi, S. Weiner, Mollusc larval shell formation: amorphous calcium carbonate is a precursor phase for aragonite. *J. Exp. Zool.* **293**, 478–491 (2002).

10. R. A. Metzler, J. A. Jones, A. J. D'Addario, E. J. Galvez, Polarimetry of Pinctada fucata nacre indicates myostracal layer interrupts nacre structure. *R. Soc. Open Sci.* **4** (2017), doi:10.1098/rsos.160893.

11. R. A. Metzler, C. Burgess, B. Regan, S. Spano, E. J. Galvez, in *The Nature of Light: Light in Nature V* (International Society for Optics and Photonics, 2014), vol. 9187, p. 918704.

12. R. A. Metzler, M. Abrecht, R. M. Olabisi, D. Ariosa, C. J. Johnson, B. H. Frazer, S. N. Coppersmith, P. U. P. A. Gilbert, Architecture of Columnar Nacre, and Implications for Its Formation Mechanism. *Phys. Rev. Lett.* **98**, 268102 (2007).

13. R. T. DeVol, R. A. Metzler, L. Kabalah-Amitai, B. Pokroy, Y. Politi, A. Gal, L. Addadi, S. Weiner, A. Fernandez-Martinez, R. Demichelis, J. D. Gale, J. Ihli, F. C. Meldrum, A. Z. Blonsky, C. E. Killian, C. B. Salling, A. T. Young, M. A. Marcus, A. Scholl, A. Doran, C. Jenkins, H. A. Bechtel, P. U. P. A. Gilbert, Oxygen Spectroscopy and Polarization-Dependent Imaging Contrast (PIC)-Mapping of Calcium Carbonate Minerals and Biominerals. *J. Phys. Chem. B*. **118**, 8449–8457 (2014).

14. P. U. P. A. Gilbert, A. Young, S. N. Coppersmith, Measurement of c-axis angular orientation in calcite (CaCO3) nanocrystals using X-ray absorption spectroscopy. *Proc. Natl. Acad. Sci.* **108**, 11350–11355 (2011).

15. H. Ade, B. Hsiao, X-ray Linear Dichroism Microscopy. *Science*. **262**, 1427–1429 (1993).

16. J. Miao, P. Charalambous, J. Kirz, D. Sayre, Extending the methodology of X-ray crystallography to allow imaging of micrometre-sized non-crystalline specimens. *Nature*. **400**, 342–344 (1999).

17. J. Miao, T. Ishikawa, I. K. Robinson, M. M. Murnane, Beyond crystallography: Diffractive imaging using coherent x-ray light sources. *Science*. **348**, 530–535 (2015).

18. Y. Shechtman, Y. C. Eldar, O. Cohen, H. N. Chapman, J. Miao, M. Segev, Phase Retrieval with Application to Optical Imaging: A contemporary overview. *IEEE Signal Process. Mag.* **32**, 87–109 (2015).

19. H. Jiang, D. Ramunno-Johnson, C. Song, B. Amirbekian, Y. Kohmura, Y. Nishino, Y. Takahashi, T. Ishikawa, J. Miao, Nanoscale Imaging of Mineral Crystals inside Biological Composite Materials Using X-Ray Diffraction Microscopy. *Phys. Rev. Lett.* **100**, 038103 (2008).

20. J. M. Rodenburg, H. M. Faulkner, A Phase Retrieval Algorithm for Shifting Illumination. *Appl. Phys. Lett.* **85**, 4795–4797 (2004).

21. P. Thibault, M. Dierolf, A. Menzel, O. Bunk, C. David, F. Pfeiffer, High-resolution scanning x-ray diffraction microscopy. *Science*. **321**, 379–382 (2008).

22. D. A. Shapiro, Y.-S. Yu, T. Tyliszczak, J. Cabana, R. Celestre, W. Chao, K. Kaznatcheev, A. L. D. Kilcoyne, F. Maia, S. Marchesini, Y. S. Meng, T. Warwick, L. L. Yang, H. A. Padmore, Chemical composition mapping with nanometre resolution by soft X-ray microscopy. *Nat. Photonics*. **8**, 765–769 (2014).

23. M. E. Birkbak, M. Guizar-Sicairos, M. Holler, H. Birkedal, Internal structure of sponge glass fiber revealed by ptychographic nanotomography. *J. Struct. Biol.* **194**, 124–128 (2016).

24. I. Zanette, B. Enders, M. Dierolf, P. Thibault, R. Gradl, A. Diaz, M. Guizar-Sicairos, A. Menzel, F. Pfeiffer, P. Zaslansky, Ptychographic X-ray nanotomography quantifies mineral distributions in human dentine. *Sci. Rep.* **5**, 9210 (2015).

25. M. Dierolf, A. Menzel, P. Thibault, P. Schneider, C. M. Kewish, R. Wepf, O. Bunk, F. Pfeiffer, Ptychographic X-ray computed tomography at the nanoscale. *Nature*. **467**, 436–439 (2010).

26. J. Deng, Y. H. Lo, M. Gallagher-Jones, S. Chen, A. Pryor, Q. Jin, Y. P. Hong, Y. S. G. Nashed, S. Vogt, J. Miao, C. Jacobsen, Correlative 3D x-ray fluorescence and ptychographic tomography of frozen-hydrated green algae. *Sci. Adv.* **4**, eaau4548 (2018).

27. P. Ferrand, A. Baroni, M. Allain, V. Chamard, Quantitative imaging of anisotropic material properties with vectorial ptychography. *Opt. Lett.* **43**, 763–766 (2018).

28. Y. H. Lo, C.-T. Liao, J. Zhou, A. Rana, C. S. Bevis, G. Gui, B. Enders, K. M. Cannon, Y.-S. Yu, R. Celestre, K. Nowrouzi, D. Shapiro, H. Kapteyn, R. Falcone, C. Bennett, M. Murnane, J. Miao, Multimodal x-ray and electron microscopy of the Allende meteorite. *Sci. Adv.* **5**, eaax3009 (2019).

29. C. Ophus, Four-Dimensional Scanning Transmission Electron Microscopy (4D-STEM): From Scanning Nanodiffraction to Ptychography and Beyond. *Microsc. Microanal.* **25**, 563–582 (2019).

30. R. Celestre, K. Nowrouzi, D. A. Shapiro, P. Denes, J. M. Joseph, A. Schmid, H. A. Padmore, Nanosurveyor 2: A Compact Instrument for Nano-Tomography at the Advanced Light Source. *J. Phys. Conf. Ser.* **849**, 012047 (2017).

31. D. A. Shapiro, R. Celestre, P. Denes, M. Farmand, J. Joseph, A. L. D. Kilcoyne, Stefano Marchesini, H. Padmore, S. V. Venkatakrishnan, T. Warwick, Y.-S. Yu, Ptychographic Imaging of Nano-Materials at the Advanced Light Source with the Nanosurveyor Instrument. *J. Phys. Conf. Ser.* **849**, 012028 (2017).

32. D. Doering, N. C. Andresen, D. Contarato, P. Denes, J. M. Joseph, P. McVittie, J. Walder, J. T. Weizeorick, B. Zheng, High speed, direct detection 1k Frame-Store CCD sensor for synchrotron radiation. *2011 IEEE Nucl. Sci. Symp. Conf. Rec.*, 1840–1845 (2011).

33. A. Maiden, D. Johnson, P. Li, Further improvements to the ptychographical iterative engine. *Optica*. **4**, 736–745 (2017).

34. M. Lerotic, R. Mak, S. Wirick, F. Meirer, C. Jacobsen, MANTiS: a program for the analysis of X-ray spectromicroscopy data. *J. Synchrotron Radiat.* **21**, 1206–1212 (2014).

35. L. Muscatine, C. Goiran, L. Land, J. Jaubert, J.-P. Cuif, D. Allemand, Stable isotopes (δ13C and δ15N) of organic matrix from coral skeleton. *Proc. Natl. Acad. Sci.* **102**, 1525–1530 (2005).

36. L. Rokach, O. Maimon, in *Data Mining and Knowledge Discovery Handbook*, O. Maimon, L. Rokach, Eds. (Springer US, Boston, MA, 2005; https://doi.org/10.1007/0-387-25465-X_15), pp. 321–352.

37. A. Pryor, Y. Yang, A. Rana, M. Gallagher-Jones, J. Zhou, Y. H. Lo, G. Melinte, W. Chiu, J. A. Rodriguez, J. Miao, GENFIRE: A generalized Fourier iterative reconstruction algorithm for high-resolution 3D imaging. *Sci. Rep.* **7**, 10409 (2017).

38. Y. Yang, C.-C. Chen, M. C. Scott, C. Ophus, R. Xu, A. Pryor, L. Wu, F. Sun, W. Theis, J. Zhou, M. Eisenbach, P. R. C. Kent, R. F. Sabirianov, H. Zeng, P. Ercius, J. Miao, Deciphering chemical order/disorder and material properties at the single-atom level. *Nature*. **542**, 75 (2017).

39. J. Zhou, Y. Yang, Y. Yang, D. S. Kim, A. Yuan, X. Tian, C. Ophus, F. Sun, A. K. Schmid, M. Nathanson, H. Heinz, Q. An, H. Zeng, P. Ercius, J. Miao, Observing crystal nucleation in four dimensions using atomic electron tomography. *Nature*. **570**, 500–503 (2019).

40. X. Tian, D. S. Kim, S. Yang, C. J. Ciccarino, Y. Gong, Y. Yang, Y. Yang, B. Duschatko, Y. Yuan, P. M. Ajayan, J. C. Idrobo, P. Narang, J. Miao, Correlating the three-dimensional atomic defects and electronic properties of two-dimensional transition metal dichalcogenides. *Nat. Mater.*, 1–7 (2020).

41. M. C. Scott, C.-C. Chen, M. Mecklenburg, C. Zhu, R. Xu, P. Ercius, U. Dahmen, B. C. Regan, J. Miao, Electron tomography at 2.4-ångström resolution. *Nature*. **483**, 444 (2012).

42. J. Miao, D. Sayre, H. N. Chapman, Phase retrieval from the magnitude of the Fourier transforms of nonperiodic objects. *JOSA A*. **15**, 1662–1669 (1998).

43. M. Farmand, R. Celestre, P. Denes, A. L. D. Kilcoyne, S. Marchesini, H. Padmore, T. Tyliszczak, T. Warwick, X. Shi, J. Lee, Y.-S. Yu, J. Cabana, J. Joseph, H. Krishnan, T. Perciano, F. R. N. C. Maia, D. A. Shapiro, Near-edge X-ray refraction fine structure microscopy. *Appl. Phys. Lett.* **110**, 063101 (2017).

44. M. R. Howells, T. Beetz, H. N. Chapman, C. Cui, J. M. Holton, C. J. Jacobsen, J. Kirz, E. Lima, S. Marchesini, H. Miao, D. Sayre, D. A. Shapiro, J. C. H. Spence, D. Starodub, An assessment of the resolution limitation due to radiation-damage in X-ray diffraction microscopy. *J. Electron Spectrosc. Relat. Phenom.* **170**, 4–12 (2009).

45. J. Marrison, L. Räty, P. Marriott, P. O'Toole, Ptychography – a label free, high-contrast imaging technique for live cells using quantitative phase information. *Sci. Rep.* **3** (2013), doi:10.1038/srep02369.

46. K. Benzerara, N. Menguy, M. Obst, J. Stolarski, M. Mazur, T. Tylisczak, G. E. Brown, A. Meibom, Study of the crystallographic architecture of corals at the nanoscale by scanning transmission X-ray microscopy and transmission electron microscopy. *Ultramicroscopy*. **111**, 1268–1275 (2011).

47. I. Coronado, M. Fine, F. R. Bosellini, J. Stolarski, Impact of ocean acidification on crystallographic vital effect of the coral skeleton. *Nat. Commun.* **10**, 1–9 (2019).

48. L. C. Nielsen, D. J. DePaolo, J. J. De Yoreo, Self-consistent ion-by-ion growth model for kinetic isotopic fractionation during calcite precipitation. *Geochim. Cosmochim. Acta*. **86**, 166–181 (2012).

49. B. R. Constantz, Coral Skeleton Construction: A Physiochemically Dominated Process. *PALAIOS*. **1**, 152–157 (1986).

50. L. Addadi, S. Weiner, Biomineralization: mineral formation by organisms. *Phys. Scr.* **89**, 098003 (2014).

51. B. H. Frazer, B. Gilbert, B. R. Sonderegger, G. De Stasio, The probing depth of total electron yield in the sub-keV range: TEY-XAS and X-PEEM. *Surf. Sci.* **537**, 161–167 (2003).

52. M. Gallagher-Jones, C. Ophus, K. C. Bustillo, D. R. Boyer, O. Panova, C. Glynn, C.-T. Zee, J. Ciston, K. C. Mancia, A. M. Minor, J. A. Rodriguez, Nanoscale mosaicity revealed in peptide microcrystals by scanning electron nanodiffraction. *Commun. Biol.* **2**, 26 (2019).

53. C. Donnelly, M. Guizar-Sicairos, V. Scagnoli, S. Gliga, M. Holler, J. Raabe, L. J. Heyderman, Three-dimensional magnetization structures revealed with X-ray vector nanotomography. *Nature*. **547**, 328–331 (2017).

54. P. Ferrand, M. Allain, V. Chamard, Ptychography in anisotropic media. *Opt. Lett.* **40**, 5144–5147 (2015).

**FIGURES**

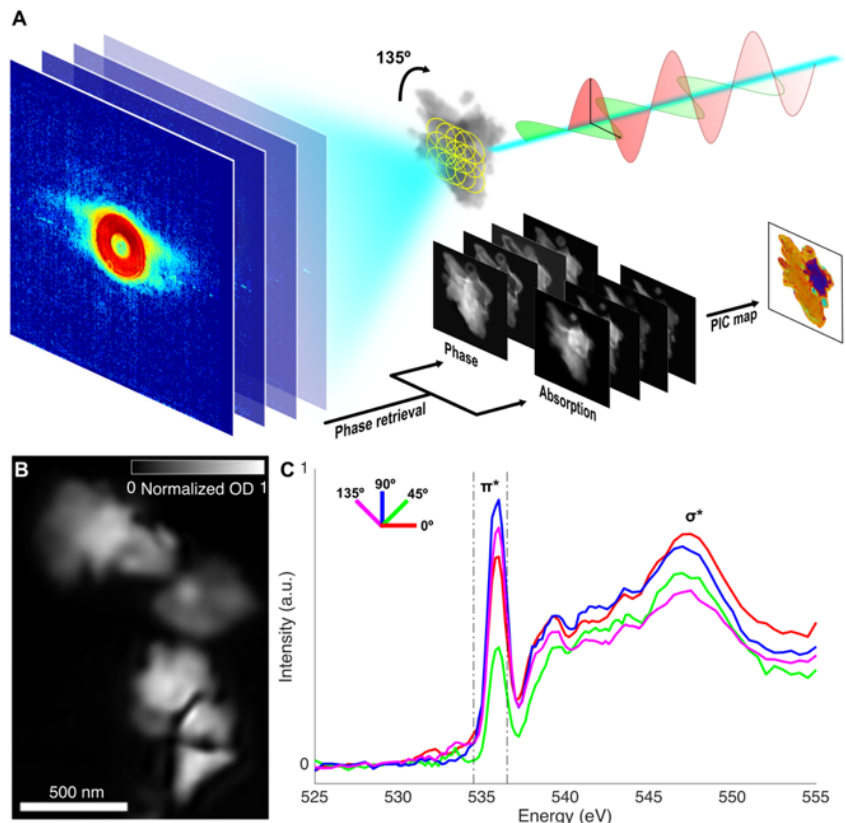

**Fig. 1. X-ray linear dichroic ptychography imaging setup.** (**A**) Experimental schematic of the x-ray linear dichroic diffraction microscope. Horizontally and vertically polarized x-rays incident on the specimen as spatially overlapping diffraction patterns are acquired at below (534.5 eV) and on (536.5 eV) the O K-edge absorption edge to obtain 0º and 90º polarization data, then sample is rotated 135º and measured again to obtain the 45º and 135º data. The diffraction patterns are then directly phased to obtain high-resolution polarization-dependent ptychography images, from which the absorption images are used to compute the PIC maps. (**B**) Ptychography absorption image of a coral particle used to collect linear dichroic absorption spectrums. (**C**) Experimental XAS spectra of the coral particle at 4 polarizations, showing a dependence of the $CaCO^3$'s $\pi^*$ peak intensity on incident x-ray polarization angles.

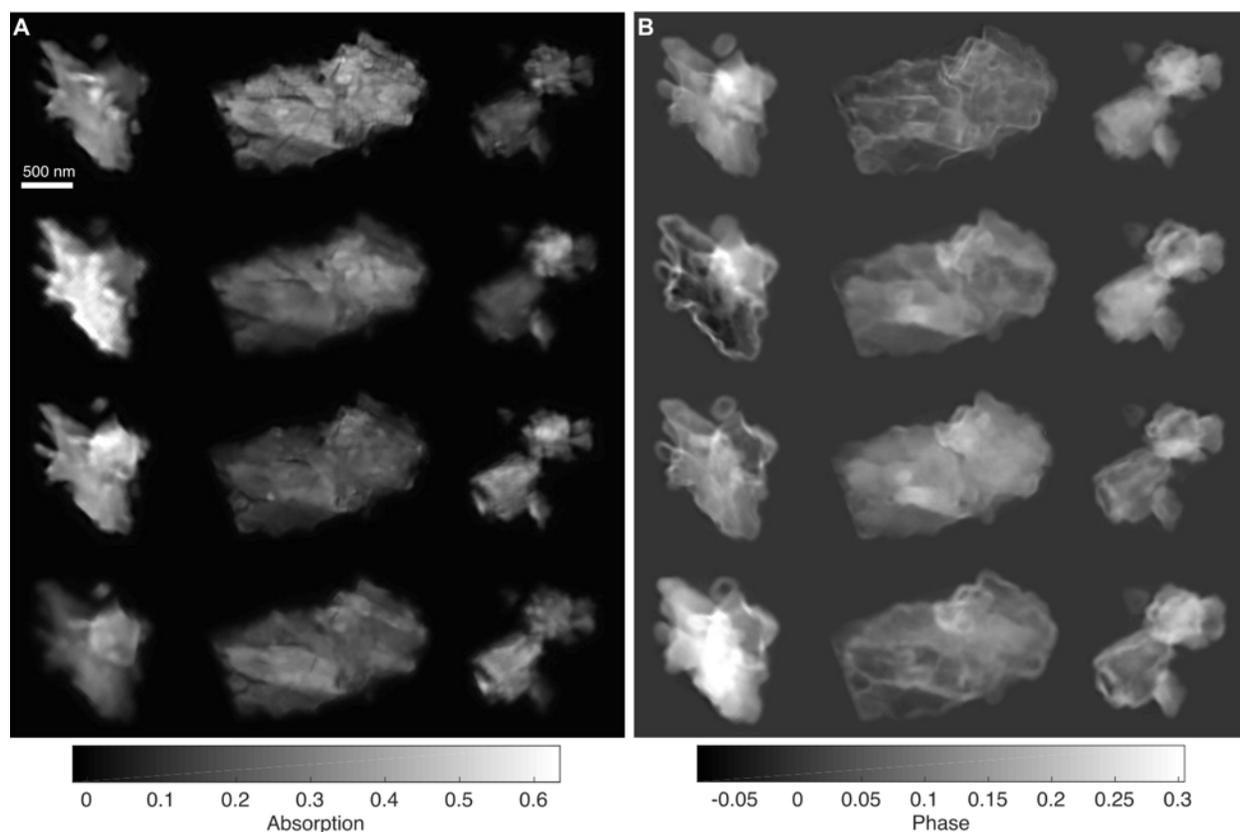

**Fig. 2. X-ray linear dichroic ptychography of coral skeleton particles.** (**A**) Ptychography absorption images of 3 aragonite particles recorded on the O K-edge absorption resonance at 536.5 eV, across 4 linear polarizations (top to bottom: 0º, 45º, 90º and 135º), showing strong polarization-dependent absorption contrast and revealing nanoscale morphologies ranging from smooth homogeneous particles several hundred nm in size to sub-100-nm fine features. (**B**) Ptychography phase images of the same particles and polarizations recorded at an energy slightly before O K-edge absorption edge of 534.5 eV, showing strong polarization-dependent phase contrast and more edge-sensitive features in internal coral structures.

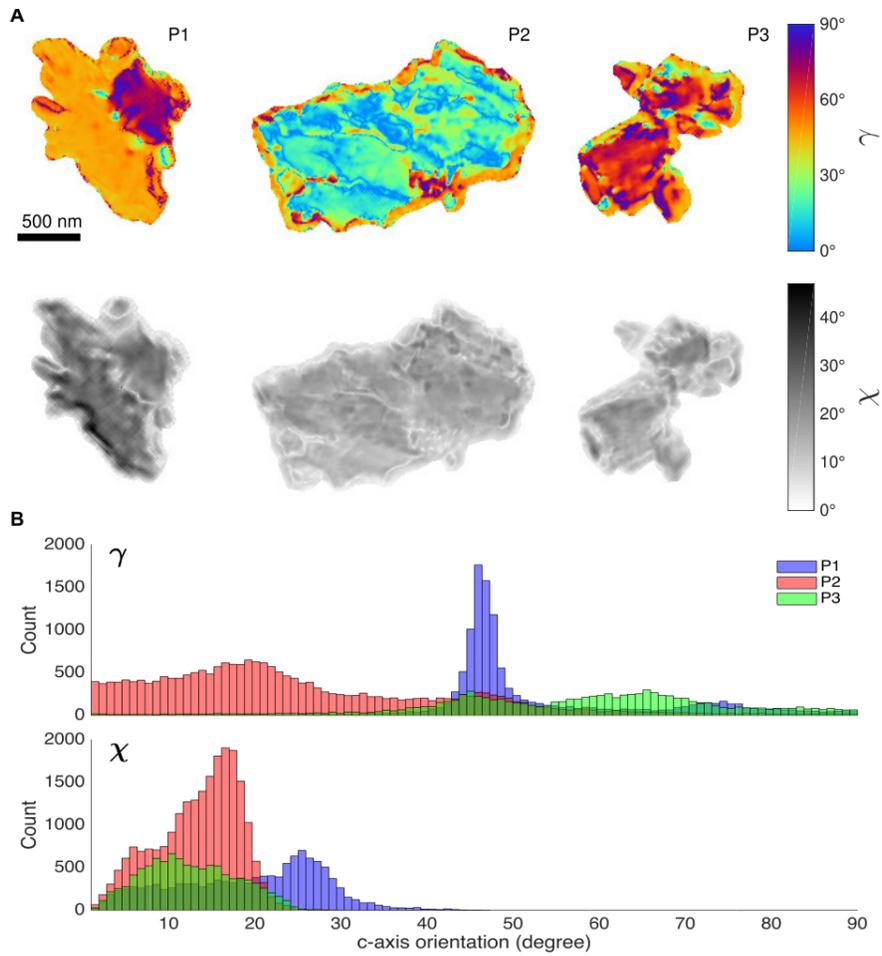

**Fig. 3. Ptychography PIC map of aragonite coral skeleton particles.** (**A**) Quantitative PIC maps of the 3 aragonite particles, calculated using 0º, 45º and 90º linear dichroic ptychography images. Hue (top row) denotes in-plane azimuthal crystal *c*-axis angle ($\gamma$) of the crystallite, while brightness (bottom row) denotes out-of-plane *c*-axis angle ($\chi$), all ranging from 0º to 90º. P1 consists of mostly homogeneous orientations, whereas P2 and P3 show much more orientational diversity. (**B**) Histograms of in-plane ($\gamma$, top) and out-of-plane ($\chi$, bottom) angles for the 3 particles, showing a narrow $\gamma$ angular spread for P1 of <35°, and much broader spread for P2 and P3 of >35°, suggesting the presence of both spherulitic and randomly oriented submicron crystallites at the nanoscopic scale.

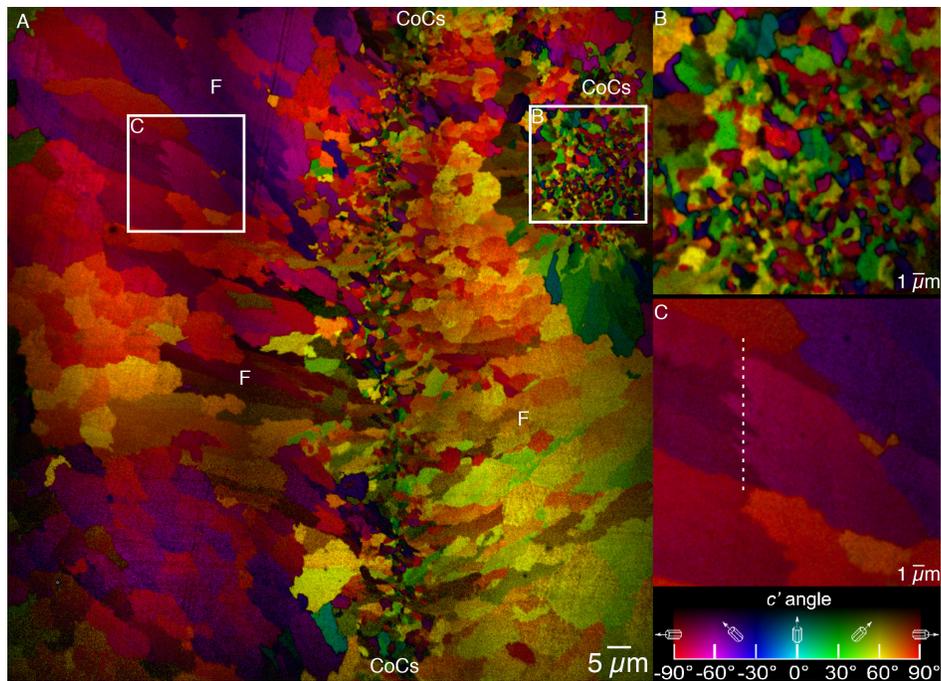

**Fig. 4. PEEM-PIC map of another sample from the same *S. aculeata* coral skeleton**. The white boxes indicate the areas magnified in panel (B) and (C). (**A**) PIC map, where color (hue and brightness) represent the in-plane and off-plane angles of the *c*-axis with respect to the polarization plane. The centers of calcification (CoCs) extend along the vertical line between the two CoCs labels, and in another area on the right (B). Spherulitic fiber (F) crystals (e.g. in C) radiate out of the CoCs, and their angular spreads are narrowly distributed, always within 35°. (**B**) The CoCs exhibit randomly oriented and thus randomly colored submicron (200-2000 nm) crystals, which have a broadly distributed angular spread. (**C**) Magnified spherulitic fiber crystals, exhibiting jagged edges. The dashed line indicates a hypothetical cut surface, resulting in spatially separate, but consistently oriented domains with two interspersed orientations.

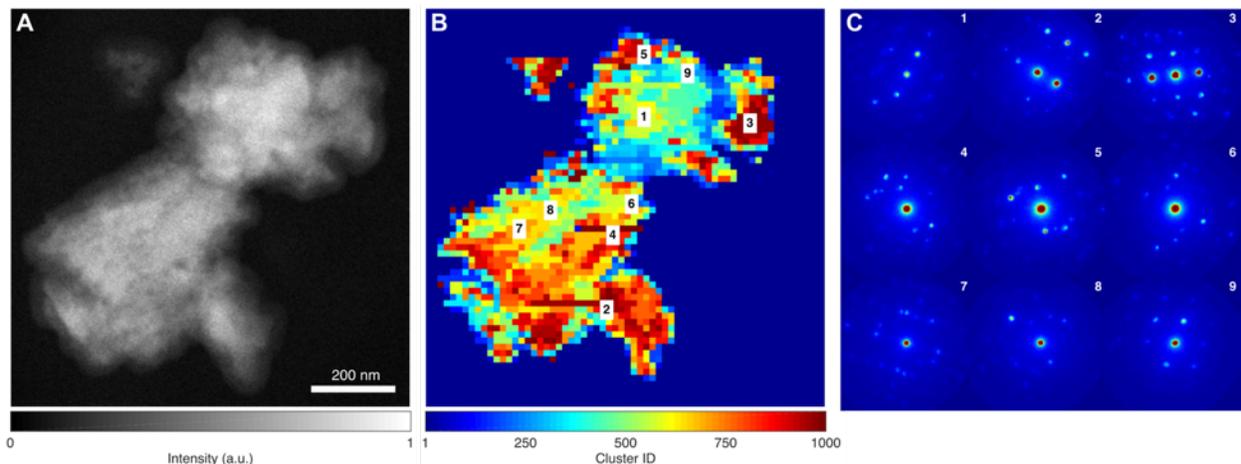

**Fig. 5. Diffraction similarity map from 4D-STEM with hierarchical clustering**. (**A**) STEM image of particle P3, which was used to acquire scanning electron nano-diffraction patterns. (**B**) Crystal axis similarity map generated using hierarchical clustering of diffraction patterns. Areas with comparable color resemble subdomains with similar crystal orientations. The resulting map

qualitatively agrees with the PIC map generated from ptychography PIC mapping (Fig. 3 P3). (**C**) Representative converging beam electron diffraction patterns from various regions of the coral particle, co-labeled in B and C, showing nanoscale orientational diversity. Scale bar: 200 nm.

## SUPPLEMENTARY MATERIALS

Fig. S1. *Seriatopora aculeata* coral skeleton.
Fig. S2. STXM-XAS spectral decomposition.
Fig. S3. X-ray linear dichroic ptychography of coral particles at other energies.
Fig. S4. Estimated ptychography resolution.
Fig. S5. Ptychography polarization-dependent contrast (PIC) map of aragonite particles with second set of polarizations.
Fig. S6. Electron tomography of P1.
Fig. S7. Monochromatic version of Fig. 3 in the main manuscript.
Fig. S8. Monochromatic version of Fig. S5 in supplementary materials.

## TEASER

X-ray linear dichroic ptychography of coral skeleton particles reveals crystal orientations with 35 nm spatial resolution.